\documentstyle[11pt,newpasp,twoside,epsf]{article}
\def\REdd{\dot M/\dot M_{Edd}}
\def\lsim{\ifmmode{\lower0.3em\hbox{$\,\buildrel <\over\sim\,$}}\else{\lower0.3em\hbox{\,\buildrel <\over\sim\,}}\fi}
\def\gsim{\ifmmode{\lower0.3em\hbox{$\,\buildrel >\over\sim\,$}}\else{\lower0.3em\hbox{\,\buildrel >\over\sim\,}}\fi}
\def\kms{\ifmmode{~{\rm km~s^{-1}}}\else{~km~s$^{-1}$}\fi}
\def\Msol{\ifmmode{~{\rm M_{\odot}}}\else{M$_{\odot}$}\fi}
\def\ion#1#2{\hbox{#1{\sc\,#2}}}

\def\edcomment#1{\iffalse\marginpar{\raggedright\sl#1\/}\else\relax\fi}
\reversemarginpar

\begin{document}
\title{Active Galaxies With Double-Peaked Emission Lines and 
What They Imply About the ``Broad-Line Region''}
\author{Michael Eracleous}
\affil{Department of Astronomy \& Astrophysics, The Pennsylvania State 
University, 525 Davey Lab, University Park, PA 16802}

\begin{abstract}

I review the distinguishing observational characteristics of active
galaxies with double-peaked emission lines and their implications for
the nature of the line-emitting region. Since double-peaked lines most
likely originate in the outer parts of the accretion disk, they can be
used to study the structure and dynamics of the disk and the
associated wind. Such studies lead to general inferences about the
broad-line regions of all AGNs. To this end, I describe the results of
recent UV spectroscopy of double-peaked emitters that probes the
disk-wind relation. I also summarize efforts to exploit the
variability of the lines to study dynamical and thermal phenomena in
the disk.

\end{abstract}

\section{Introduction: Properties of Double-Peaked Emitters and Implications}

Active galaxies with double-peaked emission lines (hereafter
double-peaked emitters) make up a small fraction of nearby ($z<0.4$)
AGNs. They are found in about 20\% of the radio-loud AGNs surveyed by
Eracleous \& Halpern (1994,2003) and in about 4\% of (radio-loud and
radio-quiet) objects from the SDSS studied by Strateva et al. (2003;
see also Strateva et al., this volume).  Double-peaked emitters share
a number of spectroscopic properties that set them apart from the
average AGN and suggest a close relation to LINERs.  These properties
include: (a) unusually-strong low-ionization emission lines from the
narrow-line region, (b) a large contribution of starlight to the
optical continuum, and (c) Balmer lines that are, on average, a few
times broader than those of other AGNs. The connection to LINERs is
bolstered by the fact that a number of double-peaked emitters have
Oxygen line ratios that satisfy the LINER definition and by the fact
that a number of previously known LINERs were recently found to host
double-peaked Balmer lines.  About 40-50\% of double-peaked H$\alpha$
profiles can be described quite well by the relativistic, circular,
Keplerian disk model of Chen, Halpern, \& Filippenko (1989) and Chen
\& Halpern (1989); two examples are shown in Figure~1. The remaining
profiles require more sophisticated models, in which the disk is not
axisymmetric (e.g., elliptical disks or disks with bright spots or
spiral arms).

The properties of double-peaked emitters can be interpreted in the
context of the scenario of Chen \& Halpern (1989) who suggested that
the inner accretion disk has the form of an ion torus (Rees et
al. 1982; known today as a radiatively inefficient accretion
flow). Such a vertically-extended structure can illuminate the
geometrically thin, outer disk and power the emission of double-peaked
lines; external illumination is needed because the line luminosity is
too high to be powered by local viscous dissipation. The same scenario
can also explain the other spectroscopic properties of disk-like
emitters, since the spectral energy distribution of an ion torus lacks
the UV bump that is a trademark of emission from an optically thick
inner disk (see discussion in Eracleous \& Halpern 1994, 2003). It is
noteworthy that several alternatives to accretion disk emission have
been proposed and discussed in the literature. However, accretion disk
emission is the interpretation favored by the data available today
(see Eracleous \& Halpern 2003 for a description of alternative
scenarios and their comparison with observations).

\begin{figure}
\plotone{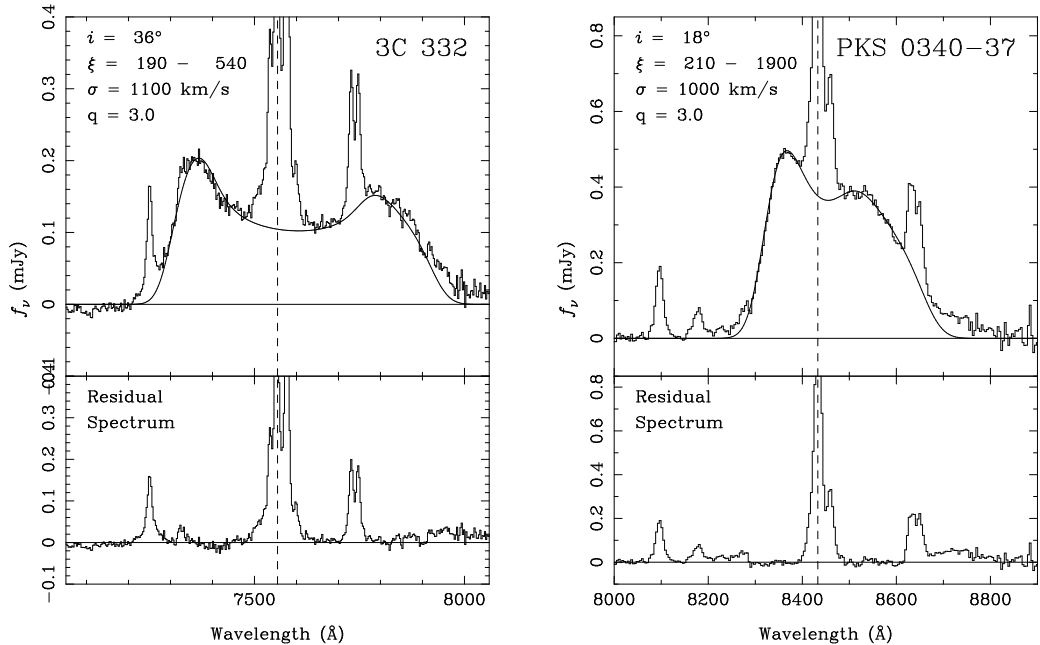}
\caption{Two examples of double-peaked H$\alpha$ profiles that can be
well fitted by a simple, relativistic, circular disk model. The top panel
shows the H$\alpha$ spectrum after continuum subtraction with the model 
superposed as a solid line. The lower panel shows the residual after 
subtraction of the model.}
\end{figure}

\section{UV Spectra of Double-Peaked Emitters and the Disk-Wind Relation}

The UV spectra of double-peaked emitters obtained with the {\it Hubble
  Space Telescope} have revealed a dramatic difference between the
profiles of the Balmer and UV lines. A good example of this difference
is provided by Arp~102B (Halpern et al. 1996). The profiles of some of
the strong optical and UV lines of Arp~102B are compared in the left
panel of Figure~2. The Balmer and \ion{Mg}{ii} lines of Arp~102B are
double-peaked with ${\rm FWHM\approx 16,000~km~s^{-1}}$, while the
far-UV lines (e.g., Ly$\alpha$, \ion{C}{iii}], \ion{C}{iv}) are single
  peaked with bell-shaped profiles that have ${\rm FWHM\approx
    3,500~km~s^{-1}}$. Moreover, once a disk model is subtracted from
  the double-peaked H$\alpha$ profile, the residual resembles the
  profiles of the UV lines. Figure~2 includes two more examples of
  double-peaked emitters displaying similar behavior: PKS~0921--213
  and NGC~1097 (a nearby LINER). The far-UV lines of PKS~0921--213 are
  single-peaked and quite strong relative to the Balmer lines,
  resembling those of typical Seyfert galaxies. On the other hand, the
  far-UV lines of NGC~1097 are fairly weak in comparison to the Balmer
  lines and they include associated absorption troughs, which are
  slightly blueshifted.  These absorption lines are another
  manifestation of a dense wind and they are particularly prominent in
  objects where the single-peaked far-UV lines are relatively weak or
  absent (see the discussion and illustrations in Eracleous, Halpern,
  \& Charlton 2003 and Eracleous 2002).

\begin{figure}
\plotone{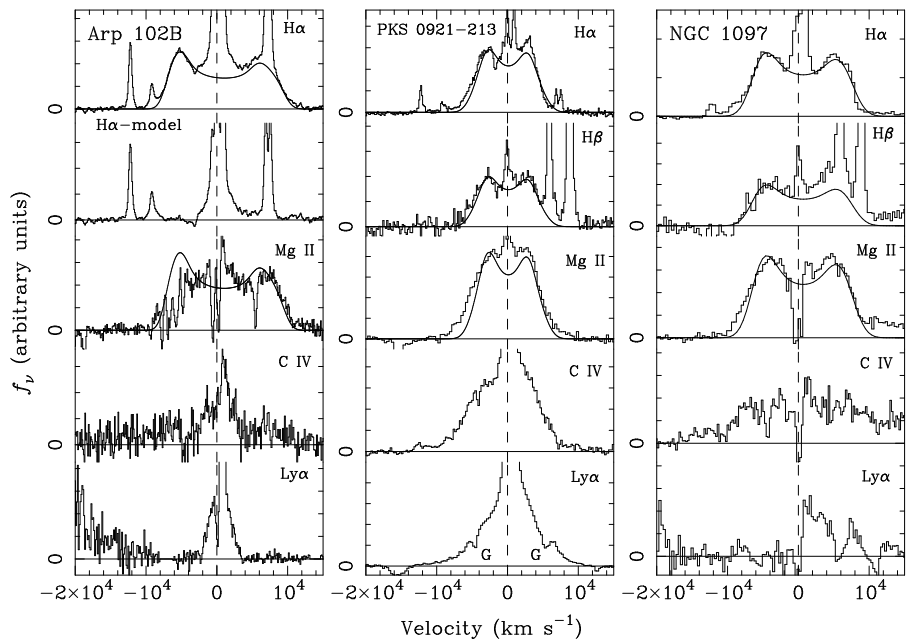}
\caption{Optical, near-UV and far-UV emission line profiles of three
  double-peaked emitters. A circular-disk model is superposed on the
  Balmer and \ion{Mg}{ii} profiles for comparison. The H$\alpha$
  profile of Arp~102B after subtraction of the model is included, to
  show that the residual resembles the Ly$\alpha$ profile. Absorption
  line marked with a ``G'' come from the gas in the Milky way, while
  all others arise in gas associated with the AGN.}
\end{figure}

The UV spectra of double-peaked emitters lead to a number of
interesting conclusions and general insights into AGN broad-line
regions as follows. There seem to be at least two line-emitting
regions in the same object, one producing the double-peaked Balmer and
\ion{Mg}{ii} lines, and another producing the single-peaked far-UV
lines. These regions can be plausibly identified with an accretion
disk consisting of dense, low-ionization gas and an outflowing wind of
higher-ionization gas. This interpretation is based on the results of
Collin-Souffrin \& Dumont (1989) and Murray \& Chiang (1997), which
show that a weakly-ionized accretion disk is a very inefficient source
of the far-UV lines; such lines should be emitted from a wind, whose
velocity structure and radiative transfer properties lead to
single-peaked line profiles. This interpretation also suggests a way
of connecting double-peaked emitters to the greater AGN population:
double-peaked emitters are the segment of the population in which the
Eddington ratio (the accretion rate relative to the Eddington rate,
$\REdd$) is very low. In this extreme the inner accretion disk turns
into an ion torus and the wind diminishes, perhaps due to a
combination of a lower mass loss rate and the harder spectral energy
distribution of the ion torus, which lacks a UV bump and makes radiative
acceleration by far-UV photons less effective (e.g., Murray et
al. 1995; Proga, Stone, \& Kallman 2000). Thus, the disk proper is
the predominant source of low-ionization lines.  Seyfert galaxies and
quasars probably represent the opposite extreme of a high Eddington
ratio, where the inner disk is geometrically thin, and the wind is the
primary source of all broad emission lines.  This scenario is
illustrated in Figure~3.

\section{Long-Term Variability of the Line Profiles and Dynamical
and Thermal Phenomena in the Disk}

Since double-peaked emission lines afford us a rare view of AGN
accretion disks, the {\it long-term} variability of their profiles can
be exploited to investigate dynamical and thermal phenomena in these
disks. However, such investigations require patience and persistence
because of the long time scales involved.  The relevant time scales
can be cast as:
\vskip -1.5em
\begin{eqnarray*}
\hbox{light-crossing time:} && \tau_{\ell c}\sim 
{r/c} \approx 6\, M_8\, \xi_3~{\rm days;} \hfil \\
\hbox{dynamical time:}      && \tau_{dyn} \sim 
(r^3/GM_{\bullet})^{1/2}\approx 6\, M_8\, \xi_3^{3/2}~{\rm months;} \\
\hbox{thermal time:}        && \tau_{th} \sim 
\tau_{dyn}/\alpha \approx 5\, (\alpha/0.1)^{-1} M_8\, \xi_3^{3/2}~{\rm years;} \\
\hbox{sound-crossing time:} && \tau_s \sim 
r/c_s \approx 70\, M_8\, \xi_3\, T_5^{-1/2}~{\rm years.} \\
\end{eqnarray*}
\vskip -1.5em \noindent
In the above expressions, $M_8$ is the black hole mass in units of
$10^8\,\Msol$, $\xi_3$ is the radial distance in the disk in units of
$10^3\, GM_{\bullet}/c^2$, $T_5$ is the temperature in units of
$10^5$~K, and $\alpha$ is the viscosity parameter of Shakura \&
Sunyaev (1973).

\begin{figure}
\plotone{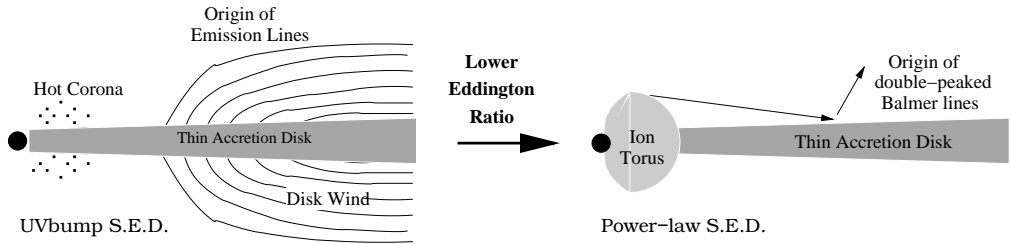}
\caption{A sketch of how the structure of the accretion disk and its
  associated wind change as the Eddington ratio goes from high values
  ($\REdd\gsim0.1$) to low values ($\REdd\lsim10^{-3}$).}
\end{figure}

One must distinguish between variability of the emission-line flux due
to reverberation of a variable continuum and variations of the
emission-line profiles caused by changes in the structure of the
line-emitting region. The former type of variation occurs on the
light-crossing time and is {\it not} accompanied by significant
profile variations, as shown by reverberation mapping of Seyfert
galaxies (e.g., Ulrich et al. 1991; Wanders \& Peterson 1996;
Kassebaum et al. 1997) and of the double-peaked emitters 3C~390.3 and
Arp~102B (Dietrich et al. 1998; Sergeev et al. 2002; Shapovalova et
al. 2000; Sergeev, Pronik, \& Sergeeva 2000). Significant changes in
the line profiles occur on much longer time scales (on the order of
the dynamical time, or longer), as shown by long-term monitoring of
some of the brighter double-peaked emitters, such as Arp~102B,
3C~390.3, and 3C~332 (e.g., Newman et al. 1997; Zheng, Veilleux, \&
Grandi 1991; Gilbert et al.  1999).

A number of scenarios have been suggested and explored as explanations
of the observed long-term variations of double-peaked line
profiles. These scenarios are inspired by theoretical models for waves
and other instabilities in the disks, which can be tested using the
long-term behavior of the line profiles. The candidate scenarios
include bright spots orbiting in the disk, precessing eccentric disks,
disks with spiral waves, and even a binary broad-line region
associated with a binary black hole.  At this time, the spiral wave
scenario appears to be the most promising: it has been successfully
applied to 3C~390.3 and 3C~332 by Gilbert et al. (1999) and to
NGC~1097 by Storchi-Bergmann et al. (2003). In the case of NGC~1097
this scenario not only explains the variability trend, but it also
leads to an estimate of the precession period that is consistent with
the black hole mass inferred from stellar kinematics (see the detailed
discussion in Storchi-Bergmann et al. 2003).

Our group has been carrying out a long-term monitoring program of many
double-peaked emitters over the past decade. Examples of our most
recent results are presented in the paper by Lewis et al. in this
volume. Our observational goals are to characterize the variability
patterns, find out if these patterns recur, and if so on what time
scale.  The most obvious trend that we observe is a modulation of the
relative heights of the two peaks on time scales of order
5--10~years. Using our long-term variability data, we have been able
to reject the binary broad-line scenario for the origin of the
double-peaked lines, based on the absence of radial velocity
variations in the two peaks (Eracleous et al. 1997). This does {\it
  not} mean that we disfavor binary black holes as such; in fact
binary black holes may be important ingredients in some of the
currently viable scenarios (e.g., the elliptical disk scenario).

In parallel to the observational effort, we have been developing
parametric descriptions of dynamical models and comparing them with
observations (see, for example, Storchi-Bergmann et al. 2003). We have
found that a detailed comparison of model line profiles with the data
allows us to select a favorite model on the basis of the variability
pattern and with the help of additional constraints, such as the black
hole mass, which sets the variability time scale in the model. Since
candidate models are falsifiable, we hope to make use of our data to
identify a universal scenario, or narrow down the possibilities at the
very least. This would represent substantial progress and could lead
to insights on the mechanisms of angular momentum transport in the
outer accretion disks of AGNs and the poorly understood causes of AGN
variability.

\acknowledgements I am grateful to Jules Halpern and Karen Lewis for
their critical reading of the manuscript and for helpful comments and
suggestions. Karen Lewis also prepared the original version of the 
cartoon shown in Figure~3.

\end{document}